# Brain Neurological Constructs: The Neuronal Computational Schemes for Resolution of Life's complexities

Jahan N Schad[*]

*Retired Lawrence Berkeley National Laboratory, UCB, 376 Tharp Drive, Moraga, CA 94556, USA*

[*]**Corresponding author:** Jahan N Schad, Retired LBNL (UCB) Scientist, 376 Tharp Drive, Moraga, Ca 94556, USA, Tel: (925) 376-4126; E-mail: Jaschadn@gmail.com

## Abstract

For complex life to evolve, a sophisticated nervous system for handling its complexities was fundamental. The demand resulted in the emergence of brain's computational facility, the neuronal network. This facet of the brain is attested solidly by its inspired scientific computational neural nets which (mathematically) resolve and solve many complex problems. The presumptive general semblance of the computational operation between the two systems allows for the inference that the process in brain's neural domain also renders complexities for solution, as sets of parametric equations, like the basic implicit algorithmic formalisms underlying the operations of the scientific neural nets. This parallel is based on the fact that such devices resolve complex problems for which no declarative logical formulation is deployed. The mathematically resolved neural net problem formalism also resembles that of many theoretically known and formulated complexities which are algorithmized, in their discretized solution domains, within the context of initial and boundary value problems for direct or iterative solution by computers. The brain neuronal net algorithmization of complexities delineate the governing Equations of life and living, solutions of which are achieved by trial-and error learning, deploying rest of the nervous system and other faculties of living beings. The computational operations of the brain delineate two mental states: consciousness and the unconscious; the aware and unaware states which describes the interactive living processes involved in charting life's path.

## Introduction

Laws of nature have rendered the realities of the world- whatever their essence may be of which animated beings are a part, and are aware of to varying degrees, as they are aware of themselves- all are subjectively perceived as the results of the occurrences of various sensory stimulations. Since the sensory data are electrochemical signals in nature, the experiences of the physical realities, are the perceptions of their reconstructions (models) in the brain; which are verified / validated by the dictum of survival, through instantaneous interaction feedbacks from the physical realities; and subject to the limits of the sensitivity of the being's sensory apparatus. The brain also discerns phenomenological complexities that relate to life and living, which it also resolves within the mental model of the physical world

Such immense tasks are now believed to be accomplished through the computational operations of the brain [1]. Progress of work in the area of artificial intelligence (AI), achieved in the span of last several decades, have led to the assumption of some measure of (general) computational operation semblance between the brain and the (brain layout inspired) scientific neural network; the underpinnings of such extrapolation have wide and far reaching implication: There are billions of degrees of freedom- a connected (discretized) nodal space of synapses- available in the computational neuronal net of the brain in comparison with those of any foreseeable scientific net computer. And this offers the possibility of breaking down, to sufficient degrees of numerical refinements, any inherent governing formulations of life environment complexities for precise (implicit mathematical) algorithmizations, needed for computations; that is, solutions or simulations.

Hence, it follows that the circumstances and the laws of nature have configured beings' computational brains, and the rest of the nervous systems, for the discernment and assimilation of the world, as well as, resolution and solution of its complexities. And this implies evolutionary embedding's of operational instructions, and solutions in its construct (as neuronal patterns), which sustain beings advanced lives, deploying its environmental interactive sensing operations, some reflections of which define consciousness.

The discussion that follows, explains the concepts of the resolution of complexities, in view of the brain's presumed computational architecture and function. And expounds the attainment of brain's solutions in the context of algorithmic initial and boundary value problem resolutions- essentially, systems of simultaneous equations, representing the governing equations of life and living, in the brain model of world realities. As such, simple unified theory that explains the essences of consciousness and the unconscious states of the brain that delineates life's path is put forward. And it settles the idea of free will- a point of contention among philosophers of the past and present - on the side of the overwhelming consensus of the causality minded scientist: Rejecting it.

The proposed theory, which is based on the presumption of human brain neuronal network computational operations and its embedded constructs- patterns organized and set by evolutionary or life span events- is to be figured out as to have "butterfly effect" on the chaotic underpinnings of the societal (collective) consciousness in the areas of education, jurisprudence, and even on thoughts about socio-political ideologies.

## Background

Occurrences of some events of the mind, such as mental disturbances, dreams, hallucinations, and so-called revelations, have always been subjects of attention, curiosity, and fascination. Dreamer medicine men of the early communities to ancient philosophers and today's well known scientists and philosophers are among many who have tried to interpret or understand these events: dreams and hallucinations are justifiably attributed to (the mostly) chaotic synthesis of past experiences stored in the subconscious (unconscious), and mental disturbances are attributed to psychological and psychiatric phenomena. Treatise of Freud (1935), and Jung (1933) and findings in various schools of psychology are parts of the more serious investigations addressing these occurrences as various attribute of consciousness and unconscious [2,3].

The less dramatic and more common incidents that people of all walks of life experience, which generally go unheeded, is the sudden emergence of solutions to pending problems, or information that they had earlier tried to recall and had forgotten about. Among the deep thinkers, whether the curious stargazer of the antiquity or researcher of the recent times, experience of a moment of revelation, an "Aha" moment, as elaborated by Roger Penrose (1990) [4], has often been reported. More important cases occur in practices of scientific research when findings happen to the mind either spontaneously, or delayed, and occasionally quite unexpectedly; some at times completely outside the realm of space-time of research environment. The process of thinking does not explain, in a direct way, the appearance of the answer or the solution. Discontinuity of the thought process, between thinking and arriving at and answer and/or solution, very clearly points to a separate process, thoroughly out of the domain of (seemingly) conscious thoughts. Discoveries (revelations), always subsequent to an "Aha" moment, fall in this category and have been attributed, by some philosophers, to emergences from an imaginary world of information [5].

The functional processes of the brain that give rise to various phenomena seemingly have not received the degrees of scrutiny they deserve. Mainly, behavioral studies and cognition research have addressed the extent of the direct role of unconscious events from the viewpoint of behavioral aspects in conducts of life. The work of the physicist Mlodinow (2011) is a valuable contribution in this regard [6-8].

Efforts in the realm of artificial intelligence to develop a measure of human-like intelligence, have been of great value in addressing the general functional processes of the mind (personalized brain): With the advent of the foundation of modern computers, around 1940, thoughts of mimicking logical thinking, through the use of such devices, have been on the minds of cognitive and neuroscientists, psychologist and linguists, and experts in other field, as well. Much advance was made early on, though bounded within the limits of the algorithmic treatment of thought processes; the presence of domains of thoughts not amenable to logic proved to be the main stumbling block to the further development of artificial intelligence. This is partly due to the context sensitive nature of logical descriptions. And fundamentally due to the inconsistency of axiom based mathematics [9], which would affect computability by mechanical computers; or in general by the universal Turing Machines (1948) [10]. Computational neural network, a construct made on the basis of the layout of neurons and their theorized function in the brain- a connectionist theory- does not suffer from such drawbacks; there is no need for precise logical structure, unlike the case of the digital computers. This approach, embodying an implicit description of a desired behavior through learning, leaving no need for declarative logical statements, which was the operational basis of the earlier artificial intelligent system has enabled the emulation of some human-like mental activities.

A reductionist's view from the biological and neurophysiological angles obviously reveals the shortcomings and the difficulties of the (mainly physical or mathematical) scientific computational neural net approach in developing human like intelligence. However, chances are that with the combination of connectionist and reductionist approaches, as perceived in the neuronal group selection (NGS) theory, proposed by Edelman (1987), a lot more progress can be made [11]. To this end, great efforts in many fields related to neurosciences continue to help develop a better understanding of the mind/brain system [12].

## Hypothesis

Understanding of many complexities (governed in "essence" by laws of Physics) availed to consciousness generally require formalized rigorous symbolic logic formulations and known conditions of the domains of influences, in order to mathematically express the related phenomena. The familiar categories of problems that deploy such formalisms are the class of the initial and boundary value problems resulting from the application of conservation principals of laws of nature (for energy, momentum and mass); common in applied sciences and engineering. With few exceptions, all such problems prove only amenable to numerical solutions requiring complex manipulations to convert the formulations to a form that can render them solvable by computers. These final forms are generally sets of parametric simultaneous equations that hold true for small values, or variations, of the state variables (rendering them linear), in a discretized space and time domain. Known initial and boundary conditions, and the assumed or known behavior of the phenomena (constitutive laws) in the solution domain, - thus all parameters defined- make the problems deterministic. Inherently the number of equations and unknowns would be equal ensuring unique and non-redundant results. Progression of the solution in increments of space, time, with defined boundary conditions, either iteratively or directly provides the proper range of results. Digital computers, requiring such logical explicit mathematical formalisms, have made the solution of very many complex problems possible.

A fundamentally different approach, for solutions of certain classes of complex problems is the computational Neural Network Method [13], which has been inspired by the conceived information processing of the brain. The neural net solution approach is well suited for a host of problems where, mainly, the complexity defies development of mathematical formalism. In this method, a presumed embedded (implicit) mathematical formalism, the initial (condition) assumptions (the neural node connection weights), and the boundary (conditions) values, which are the sensing nodes inputs, make the problem mathematically well-posed The solution approach starts with the training of the neural net that can be likened to a kind of verification/validation of its implicit computational formalism. The process requires iterative signal weighting manipulations and calculations. The calculation formalism- at least in their simplest modes- is akin to the setup of a number of parametric simultaneous equations- with varied degrees of coupling among variables-, which are solved by the physical construct of the net or, alternatively and prevalently, by subordinate digital data processing means. The mathematical formalisms (equations) of this nature are very likely the underlying inherent problem resolution scheme of the neural network. The parameters of these equations, referred to as "knowledge" values, are the values of the connections weights. Changes/modifications of the knowledge values- stored in the network as results of prior iterations or trainings- towards completion of the training, requires the deployment of a (mathematical) learning rule. The rules can perhaps be very loosely likened to constitutive laws- though somewhat arbitrary for nets- in the traditional physical problem solution domains. Again, the solution process, essentially the training of the neural net, enables such devices to provide solutions to similar classes of problems that are either very difficult or impossible, to solve otherwise.

In comparison, while the traditional mathematical (numerical) approaches rely on the availability of rigorous logical formalisms to render problems solvable, in their discretized domains, through algorithmic resolutions, the power of the scientific neural networks lie in their semblance of operating in the inherently discretized (physical or mathematical) domain of network nodes, resolving problems; without any use of a priori mathematical formulations. Progress in variants of neural net computations has added much capability to the method, discussion of which is beyond the scope of this paper.

Problem solving, in the realm of Man's conscious mind (personalized brain), begins with the onset of thinking for conceptualization and or formulation. For many cases of mentally challenging problems, the solutions are generally preceded by an "Aha" moment foretelling their manifestations. Undoubtedly this is indicative of a discontinuity in thinking, which very clearly points to separate brain backend processes, thoroughly out of the domain of (seemingly) conscious thoughts. Obviously many a problem, including the ones benefiting from common sense solutions, with no indication of thought-to-solution process, rely on brain's instantaneous accommodation of solution demands.

Brain's solution dynamics can reliably be attributed to its computational operations, especially when addressing "beings" awareness of the world; a phenomenon that is only conceivable in terms of mental models in the brain, considering the electrochemical nature of the sensory signals that it receives. Brain simulations (models) of the environment require availability of computational facilities to it. And the possibility of which can be further inferred form the, the brain inspired scientific neural net computers, and their method of problem solving. As such, from a ground level computational facility perspective, the brain is fundamentally conceptualized here in the likeness of the scientific computational neural network, with a survival perfected/tuned physio-biological construct, that has evolved, in the face of the nature's complexities that configured it- Kant [14] was on the right track to suggest presence of a structure inherent to it. From the operational end, brain is equipped as well, with a biophysically aided trial and error learning faculties: Biophysical-feedback mechanisms adjusting synaptic signals- in place of the digital processor (emulating synaptic phenomena) aided solution and learning rules- to perform calculational operations of the brain. This process determines neuronal signal weight distributions, the web patterns that correctly render neuronal web solutions of complexities.

Availability of the infinitude of a priori-learned patterns, some issued from genetic heredity, and some configured earlier during lifetime, are the additional facet of the brain's computational power, which facilitates solutions of continually posed problems due to the ever-changing life boundary conditions. Other possible facet of the brain computer, its operation details, along with the very legitimate fundamental questions from the biological reductionism points of view, will remain to be worked on and perhaps not ever to be completely answered.

In summary, life's complexities are discerned and assimilated in the brain, resolved in the nodal (synaptic) spread of brain's neuronal networks (neuronal patterns) –that is implicitly algorithmized for solution- forming in essence beings' governing Equations of Life and Living. This is much in the way of many known scientific complexities which are algorithmized - in the context of initial and boundary value problems- in their discretized domains, as sets of parametric simultaneous equations to render them solvable by computational operations, directly, or iteratively.

The brain solutions of life complexities results in a simulated model of the realities. And the involved operations delineate two mental states: Consciousness and the unconscious; the aware and unaware states which delineates the interactive living processes which chart life's path with little indication of free will.

Obviously rigorous direct testing of this hypothesis has to await much further developments in understanding of brain's functional operations. However, there are anecdotal evidences that can provide some measure of verification: Presence of antecedent events in decision-making process, measured in experimental setups, which precludes volition, is a valuable example. Regarding the claim for the absence of the free will, the principal of causality, clearly laid out in the inherent implicit underpinnings of brain/s computational operations proposed here, settles the question accordingly.

It is noted that the hypothesis, despite its philosophical undertone, is deeply rooted in robust scientific inference and logical deductions.

### Corollary (1) - Complexity resolution and consciousness

Considering the evolutionary path of life, time and time keeping would be embedded in the machinery of the cells, including those of the nervous system, as it has now been scientifically determined- cell telomeres. As such, autonomous timely execution of the system embedded instructions, be it either in the form of expressions of genetic codes, or evolutionary neural constructs (patterns), result in changes in the living form and its functionality, preparing it for sustenance and survival in the environment: all aimed at ensuring continuation of the species.

Very noticeable part of the brain' computational operations are host of varied emanations; some innate, in the likes of fight or flight; and, some such as thoughts or contemplations. The latter happens in deep disciplined inquiries, among humans, seeking complexity resolution, the Aha moments. Thought driven inquiries are all in the context of the conceptual general initial and boundary value problems, where a trial and error learning process towards final resolution is involved. What is learned of the external world (experiences), at every step of the time, while mostly enhancing the computational powers of the brain- forming neuronal circuitry pattern (structures) - prepare the "initial condition" for the next step of solution, in response to varying "boundary conditions" and procession of the time. Considering the development of the very initial conditions at the start of life (obviously differing constructs and pattern contents for each individual), followed up by its evolution as a function of varying boundary conditions (different environmental exposures), vast variability in outcomes are given; though, ironically all in an autonomous way.

Functionally, the brain's trial and error process for the proper configuration of nodal signal weights (patterns) for a solution event, is equivalent to solving of sets of simultaneous equations of a certain number of unknowns-- it being the algorithmic expression of the brain resolved complexities under consideration. In such likeness, depending on the degree of complexity, number of equations (implying engagement of varied parts of the neuronal domain, by virtue of brain plasticity), and number of unknowns would vary. The solution proceeds as parameters are appropriated by trial and error, and available knowledge is used. The output of the brain system, depending on the degree to which solution parameters are correctly ascertained, can be imperfect, perfect or redundant, which determine the state of consciousness and the ensuing (input/output) interaction with the surrounding.

**Corollary (2) - Absence of free will**

Mankind has always been coping with many, clearly out of his/her own control, natural or accidental life-influencing events; and generally, regardless of any hints of causality, has attributed them to Gods' or God's Will, and Destiny, for a rational or justification. However, outside of such events, humans have always been under the perception of having total control of their conscious mind and of having power on influencing or creating some events, and in running their lives. Obviously this is in accord with the daily practices and experiences of life when, with little thinking and concentration, one seemingly decides and chooses to perform various functions and to conduct actions. Also, the ability to engage in deep thinking and concentration, seemingly by one's volition, in order to find answers or solutions to difficult problems, strongly points to the practice of "Free Will". This is seemingly further evinced by being's vast role-play in the theater of life. However, in view of the ongoing backend brain computations, it is evident that free will is inevitably a moot subject; and, consciousness despite all its free will implications is just a mental state; the brain input/output (I/O) displayed at the mind Interface! Chance due to the randomness of the conditions is the ultimate determining factor for life and living. And, the ensuing behaviors (mental and physical), in the process of the interdependent interspecies interactions, serve the overall goal of the long-term survival of the species.

In the realms of philosophical thoughts, the role of beings in their life's path has always been subject to very serious study and discourse. Some believe in Causality on logical grounds, though still allow for some measure of responsibility for one's actions, as in determinist philosophy of compatibilism, while others rely on Morality principles that necessitate free will. The question has even been addressed by some of today's scientists: Astrophysicist Stephen Hawkins and Physicist Leonard Mlodinow [15] fundamentally reject the notion of free will, only suggesting it is more pragmatic to allow for choice in one's life. Also, Mlodinow (2011), in a recent book, comes close to attributing almost all behaviors to the subliminal actions of the unconscious [8].

We note that the Birth and the Death are the two most important out of control events of consciousness path- aside from birth gene activation and the role play of epigenetic in gene activation/deactivation in life- the in-between period in a life is also dotted with many occasions of "if I had known better…" and other admittedly out of control, and life changing events…. The theory connects the dots.

**Conclusion**

The very neuronal construct of the brain and the nervous system, an adept computational equation solver, - the characteristic which has emerged from the evolutionary processes to drive species development and survival-- is a very viable presumption, which is also highly descriptive of the mathematical nature of all complex natural phenomena. As such, brain's renditions of the sensed physical world as simulated world models, and resolutions of the complexities in it, could be considered a natural brain event: The complexities would be discerned in the brain and assimilated algorithmically, in the discretized space of neuronal nodes of the neuronal network, as sets of parametric simultaneous equations, allowing solutions, in the context of initial and boundary value problems, in the mentally simulated world model, to guide beings conducts. The fact that many complexities can be expressed algorithmically by sets of simultaneous equations is verified by the cases of many scientific and engineering complexities that get resolved in such formalism before they can be investigated through computations. Resolutions of governing laws of nature, as the equations of life and living, in the brain, continually engage it for trial and error (learning) solution operations: shaping beings abilities, defining their minds' states of Unconscious and Consciousness, and their behavioral dictums, all along the evolutionary path. Consciousness, the interactive brain display of solutions for beings entanglement with evolutionary forces, updated (for changing conditions) at every waking moment, is at the service of the overall whimsical drive of survival. Self-consciousness, "Cogito ergo sum", is the behavioral dictum of survival, driving the creatures to do, what they perceive to be the results of their thinking or volition.

The apparent chaos of nature promulgates chaos in the living: its variations, variability and multitudinous, render beings susceptible to changes around them. Nonetheless, overall long-term optimization for better fitting use of resources and adaptability with the surrounding, not necessarily in human terms, is always proceeding. Thus, it can be stated that animated beings are biological Automatons, with sophisticated body and Brain machinery, equipped with sensing devices for gathering data, continually engaged in solution of complexities of nature, generally aimed at long term survival of the species, of which being is a part.

Finally, this theory, a product of a mental happening- perhaps resolving the centuries old philosophical dilemma concerning free will- if exposed to collective consciousness can perhaps have "Butterfly effect" for humane social changes; eventually shifting the burden of individual (human) misdeeds to the collective societal shortcomings, and driving improved Jurisprudence, and even implementation of humane societal ideologies.

**Acknowledgement**

Thanks are due to Dr. M. Mokarram and K. Niazi, for their valuable help in reviewing the article.